\documentclass[conference]{IEEEtran}
\IEEEoverridecommandlockouts
\usepackage{cite}
\usepackage{amsmath,amssymb,amsfonts}
\usepackage{graphicx}
\usepackage{textcomp}
\usepackage{xcolor}

\usepackage{authblk}
\usepackage{bm}
\usepackage{algorithm, algorithmicx}
\usepackage[noend]{algpseudocode}
\usepackage{cite}  

\makeatletter
\renewcommand{\fnum@figure}{Fig. \thefigure}
\makeatother

\usepackage{mathtools}

\usepackage{float}
\usepackage[caption = false]{subfig}

\def\BibTeX{{\rm B\kern-.05em{\sc i\kern-.025em b}\kern-.08em
    T\kern-.1667em\lower.7ex\hbox{E}\kern-.125emX}}
\begin{document}

\title{
Learning Cooperative Beamforming with Edge-Update Empowered Graph Neural Networks
}

\author[1]{Yunqi Wang}
\author[2]{Yang Li}
\author[2,3]{Qingjiang Shi}
\author[1]{Yik-Chung Wu}
\affil[1]{Department of Electrical and Electronic Engineering, The University of Hong Kong, Hong Kong}
\affil[2]{Shenzhen Research Institute of Big Data, Shenzhen, China}
\affil[3]{School of Software Engineering, Tongji University, Shanghai, China}
\affil[ ]{Email: yunqi9@connect.hku.hk, liyang@sribd.cn, shiqj@tongji.edu.cn, ycwu@eee.hku.hk}

\maketitle

\begin{abstract}
Cooperative beamforming design has been recognized as an effective approach in modern wireless networks to meet the dramatically increasing demand of various wireless data traffics. It is formulated as an optimization problem in conventional approaches and solved iteratively in an instance-by-instance manner. Recently, learning-based methods have emerged with real-time implementation by approximating the mapping function from the problem instances to the corresponding solutions. Among various neural network architectures, graph neural networks (GNNs) can effectively utilize the graph topology in wireless networks to achieve better generalization ability
on unseen problem sizes. 
However, the current GNNs are only equipped with the node-update mechanism, which restricts it from modeling more complicated problems such as the cooperative beamforming design,
where the beamformers are on the graph edges of wireless networks.
To fill this gap, we propose an edge-graph-neural-network (Edge-GNN) by incorporating
an edge-update mechanism into the GNN, which learns the cooperative beamforming
on the graph edges. Simulation results show that the proposed
Edge-GNN achieves higher sum rate
with much shorter computation time than state-of-the-art approaches,
and generalizes well to different numbers of base stations and user equipments.

\end{abstract}


\section{Introduction}
As wireless data traffics dramatically increase,
the inter-cell interference becomes a fundamental limitation on the user experience,
especially for the user equipments (UEs) at the cell edge.
To mitigate the inter-cell interference, the cooperative beamforming of
multiple base stations (BSs) has been recognized as an effective approach in modern wireless networks \cite{zhang2004cochannel},
ranging from cloud radio access networks \cite{shi2015large}
to cell-free massive multiple-input and multiple-output
(MIMO) systems \cite{he2021cell}.

While the cooperative beamforming desgin plays a vital role in interference mitigation, it is mathematically a non-convex optimization problem and hence
results in large computational overheads.
For example, to maximize the sum rate of the UEs
under the transmit power constraint of each BS,
the gradient projection (GP) method~\cite{bertsekas1997nonlinear} requires a lot of iterations to reach a stationary point.
On the other hand, although the
weighted minimum mean square error (WMMSE)~\cite{shi2011iteratively} approach converges with much fewer iterations, the matrix inverse in each iteration is still computationally expensive.

To facilitate the real-time implementation,
deep learning based methods for wireless resource management
have attracted a lot of attention \cite{sun2018learning, zhu2020learning, shen2020graph},
which learns a mapping function from many problem instances to their corresponding solutions.
Once the neural network is well trained, it
can infer the solution of any new problem using simple feed-forward computations, and thus is extremely fast \cite{guo2021learning}.

Along this line of research, graph neural networks (GNNs) have shown their great potential for improving the scalability and generalization ability on unseen problem sizes \cite{shen2020graph, shen2022graph, guo2021learning}.
By modeling a wireless network as a graph, the known system parameters
and unknown beamformers to be designed
can be defined as the input features and output variables on the nodes or edges
of the graph.
The advantage of graph modeling lies in its permutation equivariance (PE) property,
which guarantees that the neural network automatically
outputs a corresponding permutation whenever the indices of any two graph nodes are exchanged.
Consequently, a large number of unnecessary permuted training samples can be discarded~\cite{shen2021neural, shen2020graph, guo2021learning}.
Moreover, since the number of trainable parameters of GNNs is independent of the graph size,
the GNNs can generalize well to different problem dimensions~\cite{shen2021ai, eisen2020optimal, li2021heterogeneous}.

However, existing GNNs are only equipped with the node-update mechanism,
restricting to solving the problems with variables defined
only on the graph nodes.
Notable examples are message passing GNN (MPGNN)~\cite{shen2020graph} and 
permutation equivariant heterogeneous GNN (PGNN)~\cite{guo2021learning}.
In particular, MPGNN was proposed for beamforming design in a particular scenario, where each transmitter serves a single receiver. Thus, each transceiver pair can be defined as a graph node,
and hence the beamformers of the transceiver pairs can be defined on the nodes.
As for PGNN, it was proposed for the downlink power allocation, where
each BS adopts a pre-designed beamformer,
so that a dedicated equivalent single-antenna channel is created for each UE.
Consequently, with each UE treated as an individual node, the power variables can also be
defined on the nodes.

While \cite{shen2020graph} and \cite{guo2021learning} are pioneering works for learning
beamforming design and power allocation,
the adopted neural network architecture prohibits its extension to the more complicated problem of cooperative beamforming design, where each BS serves multiple users and each user is also served by multiple BSs.
To fill this gap, we propose a novel GNN architecture, referred to as Edge-GNN, with both node- and edge-update mechanisms.
Compared with the existing GNNs, Edge-GNN is able to define
variables on the edges, and achieves PE property with respect to both BSs and UEs, making it the first neural network architecture applicable to the more complicated problem of cooperative beamforming design.

\section{System Model and Problem Formulation}
Consider a downlink MIMO system where $M$ BSs serve $K$ UEs cooperatively. Each BS is equipped with $N$ antennas and serves all UEs, while each UE is equipped with a single antenna and served by all BSs.
Denote the channel from BS$_m$ to UE$_k$ as $\mathbf{h}_{m,k} \in \mathbb{C}^{N}, \forall m \in \mathcal{M} \triangleq\{1,\ldots,M\},  \forall k \in \mathcal{K}\triangleq\{1,\ldots,K\}$. The beamforming vector used by BS$_m$ for serving UE$_k$ is denoted as $\mathbf{v}_{m,k} \in \mathbb{C}^{N}$. With $s_k$ denoting the symbol transmitted to UE$_k$, the received signal at UE$_k$ can be expressed as
\begin{eqnarray}\label{equation-exp3 received signal at UE k}
    y_k = \sum^M_{m=1}\mathbf{h}^H_{m,k}\mathbf{v}_{m,k} s_{k} + \sum^K_{l=1,l\neq k}\sum^M_{m=1}\mathbf{h}^H_{m,k}\mathbf{v}_{m,l} s_{l} + n_k, \nonumber\\~\forall k\in\mathcal{K},
\end{eqnarray}
where $n_k$ is the noise at UE$_k$ following $\mathcal{CN}(0,\sigma_k^2)$.
The signal-to-interference-plus-noise ratio (SINR) at UE$_k$ is
\begin{equation}
    \text{SINR}_k = \frac{\left|\sum^M_{m=1}\mathbf{h}^H_{m,k}\mathbf{v}_{m,k}\right|^2}{\sum^K_{l=1,l\neq k}\left|\sum^M_{m=1}\mathbf{h}^H_{m,k}\mathbf{v}_{m,l}\right|^2+\sigma^2_k},~\forall k\in\mathcal{K},
\label{equation-exp3 SINR}
\end{equation}
and the cooperative beamforming problem for maximizing the sum-rate of all UEs
can be formulated as
\begin{subequations}\label{equation-exp3 sumrate}
\begin{equation}\label{equation-exp3 sumrate loss}
\mathop{\max}\limits_{ \left\{\mathbf{v}_{m,k}\right\}  }~
\sum^K_{k=1} \log_2 \left(1+\text{SINR}_k\right),
\end{equation}
\begin{equation}\label{equation-exp3 sumrate cons1}
~~~~~~~~~~~~~~\text{s.t.}~\sum^K_{k=1} \left\lVert\mathbf{v}_{m,k}\right\rVert^2 \leq P_{m},~\forall m\in\mathcal{M},
\end{equation}
\end{subequations}
where $P_{m}$ denotes the maximum transmit power budget at BS$_m$.

While problem \eqref{equation-exp3 sumrate}
can be solved by conventional iterative optimization algorithms,
e.g., WMMSE, the solution has to be recomputed
once the channels $\left\{\mathbf{h}_{m,k}\right\}$ change.
This instance-by-instance approach results in large computational overheads and delays.
To facilitate the real-time implementation, we strive to train a neural network that mimics the mapping function from $\left\{P_{m}\right\}$,
$\left\{\sigma^2_k\right\}$,
and $\left\{\mathbf{h}_{m,k}\right\}$ to the optimal solution  $\left\{\mathbf{v}_{m,k}\right\}$
of problem \eqref{equation-exp3 sumrate}, so that the well-trained neural network can
infer the solution of any new problem instance with simple feed-forward computations.

To this end, we express the inputs of the neural network
$\left\{P_{m}\right\}$,
$\left\{\sigma^2_k\right\}$,
and
$\left\{\mathbf{h}_{m,k}\right\}$
more compactly as $\mathbf{f}_{\text{BS}}\triangleq \left[ P_1,\cdots,P_M \right]^T$,
$\mathbf{f}_{\text{UE}}\triangleq \left[ \sigma_1^2,\cdots,\sigma_K^2 \right]^T$,
and $\mathbf{E}\in\mathbb{C}^{M\times K\times N}$ with
$\mathbf{E}_{(m,k,:)}\triangleq\mathbf{h}_{m,k}, \forall m\in\mathcal{M},~\forall k\in\mathcal{K}$.
Similarly, the output of the neural network
$\left\{\mathbf{v}_{m,k}\right\}$ can be expressed as
$\mathbf{V}\in\mathbb{C}^{M\times K\times N}$ with
$\mathbf{V}_{(m,k,:)}\triangleq\mathbf{v}_{m,k}, \forall m\in\mathcal{M},~\forall k\in\mathcal{K}$.
Then, problem \eqref{equation-exp3 sumrate} can be rewritten by
learning a mapping function
$\phi(\cdot, \cdot,  \cdot)$ such that
\begin{subequations}\label{equation-sumrate phi}
\begin{equation}\label{equation-sumrate phi loss}
    \mathop{\max}\limits_{ \phi\left(\cdot,\cdot,\cdot\right)  }~ \sum^K_{k=1} \log_2 \left(1+\text{SINR}_k\right),
\end{equation}
\begin{equation}\label{equation-sumrate phi cons2}
    ~~~~~~~~~~~~~~~~~~~~\text{s.t.}\ \  \sum^K_{k=1} \left\lVert\mathbf{V}_{(m,k,:)}\right\rVert^2 \leq P_{m},~\forall m\in\mathcal{M},
\end{equation}
\begin{equation}\label{equation-sumrate phi cons3}
    ~~~~\mathbf{V} = \phi(\mathbf{f}_{\text{BS}}, \mathbf{f}_{\text{UE}},  \mathbf{E}).
\end{equation}
\end{subequations}

Notice that problem \eqref{equation-sumrate phi} requires an inherent PE property of $\phi(\cdot,\cdot,\cdot)$.
Specifically, if the orderings of BSs and UEs are permuted, i.e.,
the entries in $\mathbf{f}_{\text{BS}}$, $\mathbf{f}_{\text{UE}}$, and $\mathbf{E}$ are permuted,
$\phi(\cdot,\cdot,\cdot)$ should permute the output $\mathbf{V}$ accordingly.
We will show in Section III that
this PE property can be maintained by
a GNN that incorporates a properly designed edge-update mechanism.

\section{Proposed Graph Modeling and Edge-GNN}    
\begin{figure}[t!]
\centering
\includegraphics[width=0.7\linewidth]{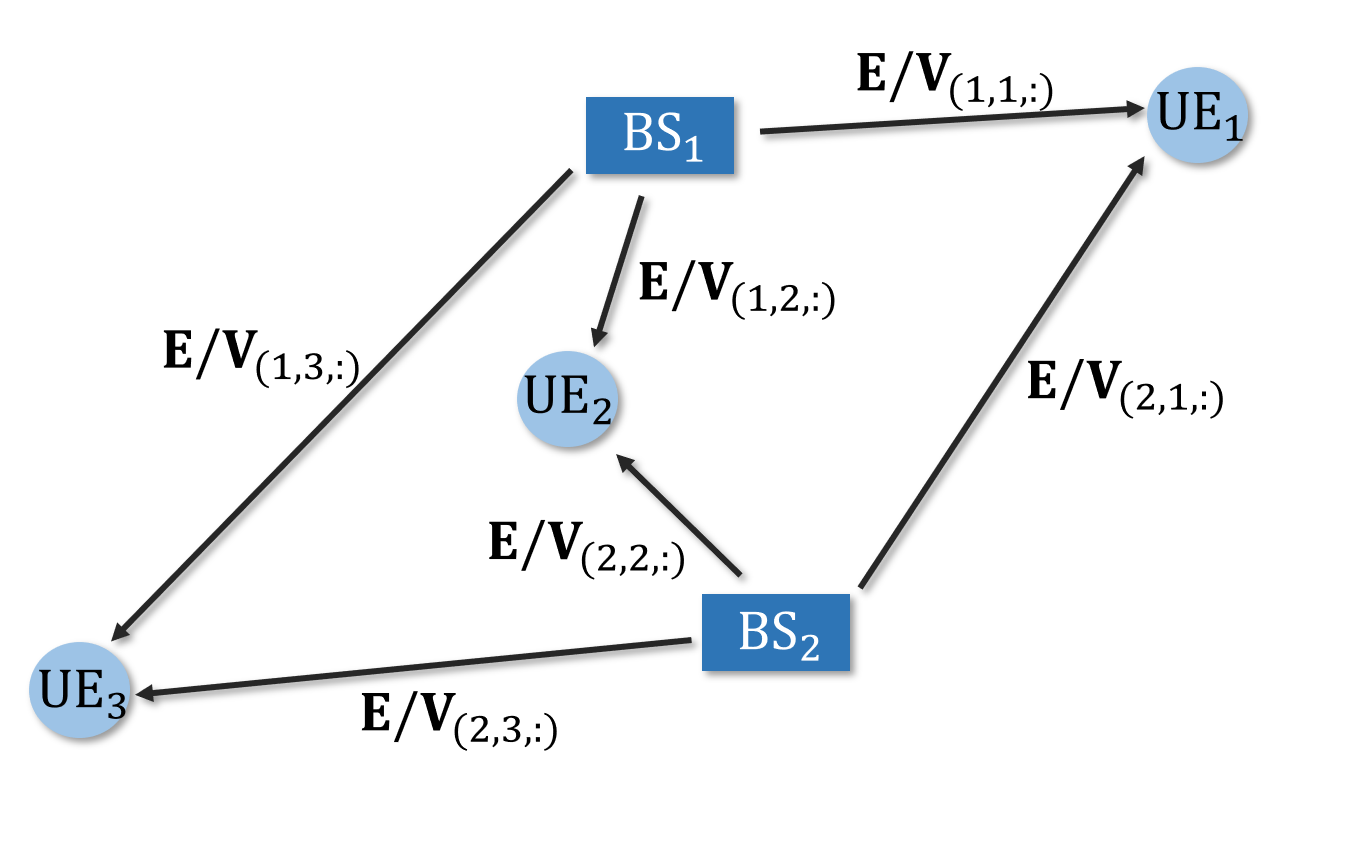}
\caption{Graph modeling of cooperative beamforming design.}
\label{fig-ex3}
\end{figure}

\subsection{Graph Modeling of Problem \eqref{equation-sumrate phi}}
Problem \eqref{equation-sumrate phi} can be modeled by a heterogeneous graph.
Specifically, the BSs and UEs can be viewed as two types of nodes, i.e., BS-nodes and UE-nodes, respectively.
Moreover, an edge is drawn between a BS-node and a UE-node if there is a communication link between them.
Such a heterogeneous graph can be expressed as $\mathcal{G} = \left\{\mathcal{M}, \mathcal{K}, \mathcal{E} \right\}$, where $\mathcal{M}$ is the set of BS-nodes,
$\mathcal{K}$ is the set of UE-nodes,
and $\mathcal{E}\triangleq\left\{(m,k)\right\}_{m\in\mathcal{M},k\in\mathcal{K}}$ is the set of edges,
respectively.

As illustrated in Fig.~\ref{fig-ex3},
by modeling the BSs and UEs as BS-nodes and UE-nodes respectively, we can incorporate the maximum transmit power budget $\mathbf{f}_{\text{BS}}$
and the noise variance $\mathbf{f}_{\text{UE}}$ as the input features on the BS-nodes and UE-nodes, respectively.
Moreover, the channel state information $\mathbf{E}$ can be viewed as the input features on the edges,
while the beamforming variable $\mathbf{V}$ can
be viewed as the outputs on the edges.

Based on the above graph modeling, we define two one-to-one
permutation mappings $\pi_{1}(\cdot)$
from $\mathcal{M}$ to $\mathcal{M}$, and $\pi_{2}(\cdot)$
from $\mathcal{K}$ to $\mathcal{K}$, respectively.
Denote a permuted problem instance of $(\mathbf{f}_{\text{BS}}, \mathbf{f}_{\text{UE}}, \mathbf{E})$ as
$(\mathbf{\dot{f}}_{\text{BS}}, \mathbf{\dot{f}}_{\text{UE}},  \mathbf{\dot{E}})$,
whose entries satisfy
$\dot{P}_{\pi_{1}(m)} = P_{m}$,
$\dot{\sigma}^2_{\pi_{2}(k)} = \sigma^2_{k}$,
and
$\mathbf{\dot{h}}_{(\pi_{1}(m),\pi_{2}(k))} = \mathbf{{h}}_{(m,k)}$,
$\forall m\in\mathcal{M},~\forall k\in\mathcal{K}$.
Denote $\mathbf{\dot{V}}=\phi(\mathbf{\dot{f}}_{\text{BS}}, \mathbf{\dot{f}}_{\text{UE}},  \mathbf{\dot{E}})$ and $\mathbf{V}=\phi(\mathbf{f}_{\text{BS}}, \mathbf{f}_{\text{UE}},  \mathbf{E})$
as the corresponding outputs of the mapping function $\phi(\cdot, \cdot,  \cdot)$, respectively.
Since $(\mathbf{\dot{f}}_{\text{BS}}, \mathbf{\dot{f}}_{\text{UE}},  \mathbf{\dot{E}})$
is just a re-ordering of the BSs and UEs in $(\mathbf{f}_{\text{BS}}, \mathbf{f}_{\text{UE}},  \mathbf{E})$, the corresponding outputs of the mapping function $\phi(\cdot, \cdot,  \cdot)$ should satisfy
\begin{equation}\label{equation-pe goal}
\mathbf{\dot{V}}_{(\pi_{1}(m),\pi_{2}(k),:)} = \mathbf{V}_{(m,k,:)},~\forall (m,k)\in\mathcal{E}.
\end{equation}
We will show in Section III-E that \eqref{equation-pe goal} can be guaranteed by
the proposed Edge-GNN with a properly designed edge-update mechanism.

\subsection{Overall Architecture of Edge-GNN}
\begin{figure}[t!]
\centering
\includegraphics[width=\linewidth]{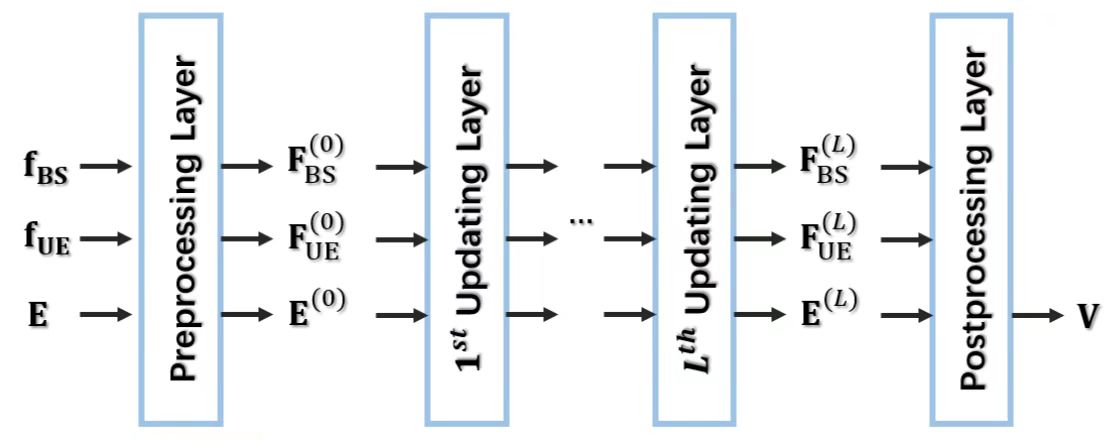}
\caption{The overall architecture of the proposed Edge-GNN, which contains a preprocessing layer, $L$ updating layers, and a postprocessing layer.}
\label{fig-network architecture}
\end{figure}

Based on the above graph modeling, we next
present the proposed Edge-GNN, which
consists of a preprocessing layer, $L$ updating layers, and a postprocessing layer as illustrated in Fig.~\ref{fig-network architecture}.

Specifically, the preprocessing layer
transforms the inputs $\left(\mathbf{f}^{\text{BS}}, \mathbf{f}^{\text{UE}}, \mathbf{E}\right)$
into the initial node and edge representations $\left(\mathbf{F}_{\text{BS}}^{(0)}\in\mathbb{R}^{M\times {d}_{\text{BS}}}, \mathbf{F}_{\text{UE}}^{(0)}\in\mathbb{R}^{K\times {d}_{\text{UE}}},  \mathbf{E}^{(0)}\in\mathbb{R}^{M\times K\times {d}_{\text{E}}}\right)$
using three node/edge-wise multi-layer perceptrons (MLPs),
where ${d}_{\text{BS}}$, ${d}_{\text{UE}}$, and ${d}_{\text{E}}$
denote the dimensions of the representations on BS-nodes, UE-nodes, and edges, respectively.
These representations are updated according to node- and edge-update mechanisms in the $L$ updating layers, where
the $l$-th updating layer takes
$\left(\mathbf{F}_{\text{BS}}^{(l-1)}, \mathbf{F}_{\text{UE}}^{(l-1)},  \mathbf{E}^{(l-1)}\right)$ as the inputs, and then outputs the updated representations $\left(\mathbf{F}_{\text{BS}}^{(l)}, \mathbf{F}_{\text{UE}}^{(l)}, \mathbf{E}^{(l)}\right)$. The dimensions of the representations will not change in the updating layers.
Finally, the postprocessing layer transforms $\mathbf{E}^{(L)}$ into the final output
$\mathbf{V}$ using an edge-wise MLP.
The magnitude of each $\mathbf{V}_{(m,k,:)}$ is
normalized to satisfy the maximum transmit power constraint \eqref{equation-sumrate phi cons2}.

\subsection{Node-Update Mechanism}
\label{sec: node-update mechanism}
To update the node representations,
the $l$-th updating layer takes $\left(\mathbf{F}_{\text{BS}}^{(l-1)}, \mathbf{F}_{\text{UE}}^{(l-1)}, \mathbf{E}^{(l-1)}\right)$ as the inputs, and then outputs the updated node representations $\left(\mathbf{F}_{\text{BS}}^{(l)}, \mathbf{F}_{\text{UE}}^{(l)}\right)$.
In particular, when updating the representation of BS$_m$,
the inputs are composed of the previous layer's representations of
BS$_m$, and the representations of the neighbouring UEs and edges, i.e.,
UE$_{k}$ and edges $(m,k)$, $\forall k\in\mathcal{N}^{\text{BS}}_m$, where $\mathcal{N}^{\text{BS}}_m$
denotes the set of neighboring UEs of BS$_m$\footnote{For the problem of cooperative beamforming design, if BS$_m$ severs all the UEs, the set of neighboring UEs is just $\mathcal{N}^{\text{BS}}_m=\mathcal{K}$. The notation $\mathcal{N}^{\text{BS}}_m$
gives more flexibility to extend to the more general scenarios, where each BS only serves a part of UEs.}.
First, the previous layer's representations of UE$_{k}$ and edge $(m,k)$
are concatenated and then processed by an MLP.
Next, the processing results from all UE$_k$ with $k\in\mathcal{N}^{\text{BS}}_m$
are combined by an aggregation function (e.g., mean or max aggregator), which extracts the information from all the neighboring UEs regardless of their index order.
Finally, the previous layer's representation of BS$_m$ and the aggregated result are concatenated, and then
processed by another MLP.
The above procedure gives the following BS-update mechanism in the $l$-th updating layer:
\begin{equation} \label{equation-EGNN encoding BS}
\begin{split}
    \mathbf{f}_{\text{BS},m}^{(l)} = \text{MLP}_2^{(l)}\left( \mathbf{f}_{\text{BS},m}^{(l-1)} , \text{AGG}_{\text{BS}}^{(l)}  \left\{ \text{MLP}_1^{(l)}\left(\mathbf{f}_{\text{UE},k}^{(l-1)}, \right.\right.\right. \\ \left.\left.\left. \mathbf{e}_{m,k}^{(l-1)}\right)  \right\}_{k \in \mathcal{N}^{\text{BS}}_m }\right),~\forall m\in\mathcal{M},
\end{split}
\end{equation}
where $\mathbf{f}_{\text{BS},m}^{(l-1)}$ is the $m$-th row of $\mathbf{F}_{\text{BS}}^{(l-1)}$,
$\mathbf{f}_{\text{UE},k}^{(l-1)}$ is the $k$-th row of $\mathbf{F}_{\text{UE}}^{(l-1)}$,
$\mathbf{e}_{m,k}^{(l-1)}$  is the $(m,k)$-th fiber of the tensor $\mathbf{E}^{(l-1)}$,
$\text{MLP}_1^{(l)}$ and $\text{MLP}_2^{(l)}$ are two MLPs, and $\text{AGG}_{\text{BS}}^{(l)}$ is an aggregation function.

Similarly,  the UE-update mechanism in the $l$-th updating layer reverses
the roles of BSs and UEs in \eqref{equation-EGNN encoding BS}:
\begin{equation}\label{equation-EGNN encoding UE}
\begin{split}
    \mathbf{f}_{\text{UE},k}^{(l)} = \text{MLP}_4^{(l)}\left( \mathbf{f}_{\text{UE},k}^{(l-1)} , \text{AGG}_{\text{UE}}^{(l)}  \left\{ \text{MLP}_3^{(l)}\left(\mathbf{f}_{\text{BS},m}^{(l-1)}, \right.\right.\right. \\ \left.\left.\left. \mathbf{e}_{m,k}^{(l-1)}\right)  \right\}_{m \in \mathcal{N}^{\text{UE}}_k }\right),~\forall k\in\mathcal{K},
\end{split}
\end{equation}
where $\mathcal{N}^{\text{UE}}_k$ is the set of neighboring BSs of UE$_k$,
$\text{MLP}_3^{(l)}$ and $\text{MLP}_4^{(l)}$ are two MLPs, and $\text{AGG}_{\text{UE}}^{(l)}$ is an aggregation function.

Notice that the representations of BSs and UEs are updated differently in the proposed Edge-GNN.
This is different from the previous work MPGNN \cite{shen2020graph},
in which the node representations are updated homogeneously.
Moreover, in the proposed Edge-GNN, the input edge representation
$\mathbf{e}_{m,k}^{(l-1)}$ in \eqref{equation-EGNN encoding BS} and \eqref{equation-EGNN encoding UE}
are with superscript $l-1$ and hence are also
updated (see the edge-update mechanism).
Taking a similar analysis in \cite{shen2020graph},
we can conclude that the node-update mechanisms in
\eqref{equation-EGNN encoding BS} and \eqref{equation-EGNN encoding UE}
satisfy the following PE property.

$\textbf{Property~1 (PE in Node-Update Mechanism):}$ \emph{The node-update mechanisms~\eqref{equation-EGNN encoding BS} and \eqref{equation-EGNN encoding UE} are permutation equivariant with respect to BSs and UEs, respectively.
Specifically, for any permutations $\pi_1(\cdot)$ and $\pi_2(\cdot)$, we have }
\begin{subequations}
\begin{equation}\label{PE-BS}
\begin{split}
\mathbf{f}_{\text{BS},\pi_1(m)}^{(l)} = \text{MLP}_2^{(l)}\left( \mathbf{f}_{\text{BS},\pi_1(m)}^{(l-1)} , \text{AGG}_{\text{BS}}^{(l)}  \left\{ \text{MLP}_1^{(l)}\left(\mathbf{f}_{\text{UE},k}^{(l-1)}, \right.\right.\right. \\ \left.\left.\left. \mathbf{e}_{\pi_1(m),k}^{(l-1)}\right)  \right\}_{k \in \mathcal{N}^{\text{BS}}_{\pi_1(m)} }\right),~\forall m\in\mathcal{M},
\end{split}
\end{equation}
\begin{equation}\label{PE-UE}
\begin{split}
\mathbf{f}_{\text{UE},\pi_2(k)}^{(l)} = \text{MLP}_4^{(l)}\left( \mathbf{f}_{\text{UE},\pi_2(k)}^{(l-1)} , \text{AGG}_{\text{UE}}^{(l)}  \left\{ \text{MLP}_3^{(l)}\left(\mathbf{f}_{\text{BS},m}^{(l-1)}, \right.\right.\right. \\ \left.\left.\left. \mathbf{e}_{m,\pi_2(k)}^{(l-1)}\right)  \right\}_{m \in \mathcal{N}^{\text{UE}}_{\pi_2(k)} }\right),~\forall k\in\mathcal{K}.
\end{split}
\end{equation}
\end{subequations}

\subsection{Edge-Update Mechanism}
\label{sec: edge-update mechanism}
\begin{figure}[t!]
\centering
\includegraphics[width=\linewidth]{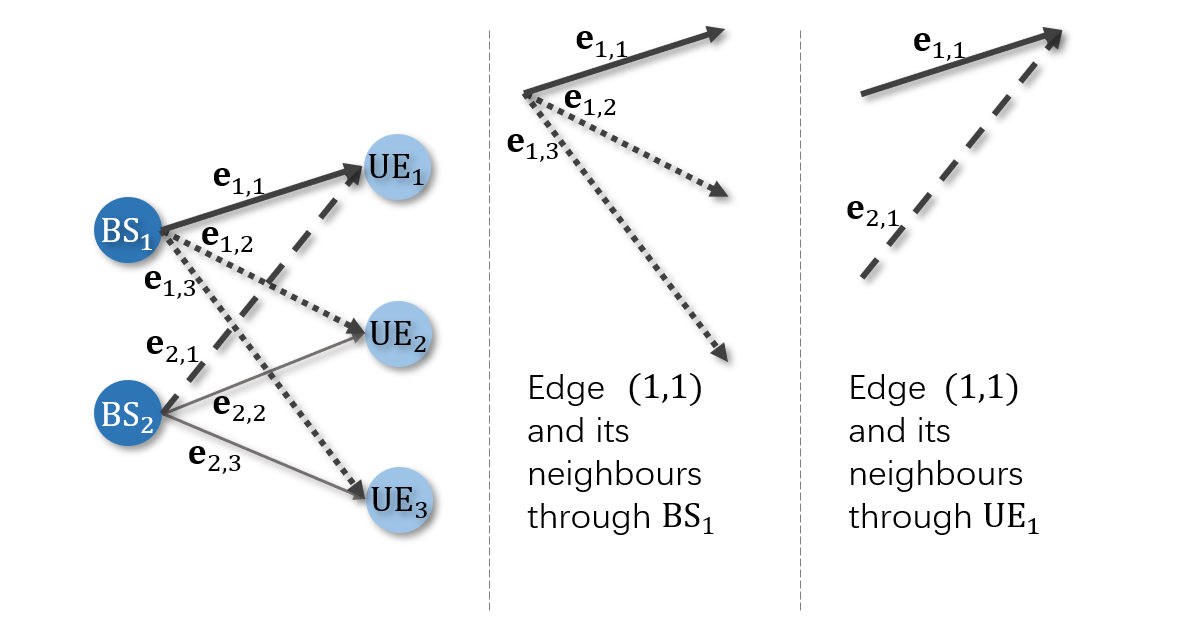}
\caption{The neighbors that share the connection with edge $(1,1)$ through $\text{BS}_1$ are edge $(1,2)$ and edge $(1,3)$, which are denoted by dotted lines. The neighbor that shares the connection with edge $(1,1)$ through $\text{UE}_1$ is edge $(2,1)$, which is denoted by a dashed line.}
\label{fig-edge neighbor}
\end{figure}

To update the edge representations,
the $l$-th updating layer takes $\left(\mathbf{F}_{\text{BS}}^{(l-1)}, \mathbf{F}_{\text{UE}}^{(l-1)}, \mathbf{E}^{(l-1)}\right)$ as the inputs, and then outputs the updated edge representations $\mathbf{E}^{(l)}$.
Different from the node-update mechanism, where
the neighbors of a BS (or UE) are clearly defined as the connecting UEs (or BSs),
it is more complicated to define the neighbors of an edge, let alone how to aggregate
their representations.
Notice that an edge may connect with other edges through
either a BS or a UE.
In particular, for the edge $(m,k)$,
its neighboring edges through BS$_m$ are $(m,k_1), \forall k_1 \in \mathcal{N}^{\text{BS}}_m \setminus \{k\}$,
while the neighboring edges through UE$_k$ are
$(m_1,k), \forall m_1 \in \mathcal{N}^{\text{UE}}_k \setminus \{m\}$.
For example, as shown in Fig.~\ref{fig-edge neighbor},
the neighbors of edge $(1,1)$ through $\text{BS}_1$ are edge $(1,2)$ and edge $(1,3)$,
whereas the neighbor of edge $(1,1)$ through $\text{UE}_1$ is edge $(2,1)$.
This causes the neighbors of an edge to be innately divided into two categories according to the connecting nodes.
Consequently, different from the node-update mechanisms in \eqref{equation-EGNN encoding BS} and~\eqref{equation-EGNN encoding UE},
the edge-update mechanism should provide two different aggregations for the two types of neighbors.

Specifically, when updating the representation of edge $(m,k)$,
the inputs are composed of the previous layer's representations of
edge $(m,k)$, BS$_{m}$, UE$_{k}$, the neighboring edges
$(m,k_1), \forall k_1 \in \mathcal{N}^{\text{BS}}_m \setminus \{k\}$,
and the neighboring edges
$(m_1,k), \forall m_1 \in \mathcal{N}^{\text{UE}}_k \setminus \{m\}$.
First, the previous layer's representations of neighboring edges $(m,k_1), \forall k_1 \in \mathcal{N}^{\text{BS}}_m \setminus \{k\}$ and the connecting node BS$_{m}$ are concatenated and then processed by an MLP. Meanwhile, the previous layer's representations of neighboring edges $(m_1,k), \forall m_1 \in \mathcal{N}^{\text{UE}}_k \setminus \{m\}$ and the connecting node UE$_{k}$ are concatenated and then processed by another MLP. Next, the above results are combined by an aggregation function, which extracts the information from all the neighboring edges regardless of their index orders.
Finally, the previous layer's representation of edge $(m,k)$ is concatenated with the above aggregated result, and then processed by an additional MLP to produce the updated representation of edge $(m,k)$.
We can express the above edge-update procedure in the $l$-th updating layer as
\begin{equation}
\begin{aligned}
\mathbf{e}_{m,k}^{(l)} =~\text{MLP}_7^{(l)}\left( \mathbf{e}_{m,k}^{(l-1)} , \text{AGG}_{\text{E}}^{(l)}  \left\{ \text{MLP}_5^{(l)}\left(\mathbf{e}_{m,k_1}^{(l-1)}, \mathbf{f}_{\text{BS},m}^{(l-1)}\right) , \right.\right.  \\
\left.\left. \text{MLP}_6^{(l)}\left(\mathbf{e}_{m_1,k}^{(l-1)}, \mathbf{f}_{\text{UE},k}^{(l-1)}\right)  \right\}_{k_1 \in \mathcal{N}^{\text{BS}}_m \setminus \{k\}, m_1 \in \mathcal{N}^{\text{UE}}_k \setminus \{m\}}\right), \\  \forall (m,k)\in\mathcal{E},
\label{equation-EGNN encoding edge}
\end{aligned}
\end{equation}
where $\text{MLP}_5^{(l)}$, $\text{MLP}_6^{(l)}$, and $\text{MLP}_7^{(l)}$ are three MLPs,
and $\text{AGG}_{\text{E}}^{(l)}$ is an aggregation function.

Compared with the node-update mechanisms~\eqref{equation-EGNN encoding BS} and \eqref{equation-EGNN encoding UE}, the edge-update mechanism~\eqref{equation-EGNN encoding edge} is more complicated.
Since the definition of neighbors in the edge-update mechanism is
more complex than that in the node-update mechanism.
In particular, the edge-update mechanism
faces a more complicated situation where the neighboring edges are innately divided into two categories according to the two possible connecting nodes.
Consequently, different from the node-update mechanisms~\eqref{equation-EGNN encoding BS} and \eqref{equation-EGNN encoding UE}, where the information from the neighbors are gathered by a single MLP,
the proposed edge-update mechanism applies
two different transformations to extract the information from two
different types of neighboring edges.
We next show that the edge-update mechanism \eqref{equation-EGNN encoding edge} enjoys the following PE property:

$\textbf{Property~2 (PE in Edge-Update Mechanism):}$ \emph{The edge-update mechanism~\eqref{equation-EGNN encoding edge} is permutation equivariant with respect to BSs and UEs.
Specifically, for any permutations $\pi_1(\cdot)$ and $\pi_2(\cdot)$, we have }
\begin{equation}  \label{equation-PE-edge}
\begin{aligned}
    \mathbf{e}_{\pi_1(m),\pi_2(k)}^{(l)} =~&\text{MLP}_7^{(l)}\left( \mathbf{e}_{\pi_1(m),\pi_2(k)}^{(l-1)} , \text{AGG}_{\text{E}}^{(l)}  \Big\{ \right. \\
     &\text{MLP}_5^{(l)}\left(\mathbf{e}_{\pi_1(m),k_1}^{(l-1)},  \mathbf{f}_{\text{BS},\pi_1(m)}^{(l-1)}\right), \\ &\text{MLP}_6^{(l)}\left(\mathbf{e}_{m_1,\pi_2(k)}^{(l-1)}, \mathbf{f}_{\text{UE},\pi_2(k)}^{(l-1)}   \right)
     \\ &\left. \Big\}_{k_1 \in \mathcal{N}^{\text{BS}}_{\pi_1(m)} \setminus \left\{\pi_2(k) \right\}, m_1 \in \mathcal{N}^{\text{UE}}_{\pi_2(k)} \setminus \left\{\pi_1(m)\right\}}\right),  \\  &~\forall (m,k)\in\mathcal{E}.
\end{aligned}
\end{equation}

\textit{Proof:} See Appendix A.

\subsection{Key Insights}
The proposed Edge-GNN for representing
$\phi(\cdot, \cdot,  \cdot)$ has been specified as
a preprocessing layer,
$L$ updating layers, and a postprocessing layer, where the
preprocessing and postprocessing layers utilize
node/edge-wise MLPs, and the $L$ updating layers are built on node- and edge-update mechanisms
\eqref{equation-EGNN encoding BS}, \eqref{equation-EGNN encoding UE}, and \eqref{equation-EGNN encoding edge}. Next,
we provide some key insights of the proposed Edge-GNN
for learning the cooperative beamforming as follows.

\subsubsection{Permutation Equivariant with Respect to BSs and UEs}
The proposed Edge-GNN enjoys the following PE property.

$\textbf{Proposition 1 (PE in Edge-GNN):}$
\emph{The proposed Edge-GNN is permutation equivariant with respect to BSs and UEs.
Specifically,
for any permutations $\pi_{1}(\cdot)$ and $\pi_{2}(\cdot)$, denote a permuted problem instance of $(\mathbf{f}_{\text{BS}}, \mathbf{f}_{\text{UE}}, \mathbf{E})$ as
$(\mathbf{\dot{f}}_{\text{BS}}, \mathbf{\dot{f}}_{\text{UE}},  \mathbf{\dot{E}})$,
whose entries satisfy
$\dot{P}_{\pi_{1}(m)} = P_{m}$,
$\dot{\sigma}^2_{\pi_{2}(k)} = \sigma^2_{k}$,
and
$\mathbf{\dot{h}}_{(\pi_{1}(m),\pi_{2}(k))} = \mathbf{{h}}_{(m,k)}$,
$\forall m\in\mathcal{M},~\forall k\in\mathcal{K}$.
Denote $\mathbf{\dot{V}}=\phi(\mathbf{\dot{f}}_{\text{BS}}, \mathbf{\dot{f}}_{\text{UE}},  \mathbf{\dot{E}})$ and $\mathbf{V}=\phi(\mathbf{f}_{\text{BS}}, \mathbf{f}_{\text{UE}},  \mathbf{E})$, respectively.
The corresponding outputs of
the proposed Edge-GNN for representing
$\phi(\cdot, \cdot,  \cdot)$ always satisfy \eqref{equation-pe goal}.}

\textit{Proof:} See Appendix B.

Proposition 1 implies that the proposed Edge-GNN is
inherently incorporated with the PE
property. This is in sharp contrast to the generic MLPs,
which require all permutations of each training sample
to approximate this property. Thus, the proposed
Edge-GNN can reduce the sample complexity and
training difficulty.

\subsubsection{Generalization on Different Numbers of BSs and UEs}
In all the layers of the proposed Edge-GNN, the dimensions of the trainable parameters
are independent of the numbers of BSs and UEs.
This scale adaptability empowers Edge-GNN
to be trained in a setup with a small number of BSs or UEs,
while being deployed to a much larger wireless network for the inference.

\subsubsection{Tackling Edge Variables}
The proposed Edge-GNN is equipped with an edge-update mechanism,
which facilitates the update of the variables on graph edges.
This allows Edge-GNN to be applied
in a wider range of scenarios.

\section{Simulation Results}
\subsection{Simulation Setting}
In this section, we demonstrate the superiority of the proposed Edge-GNN for the cooperative beamforming design via simulations.
We consider a downlink wireless network
in a $2 \times 2$ km$^2$ area, where the BSs and UEs are uniformly distributed.
Each BS is equipped with $2$ antennas and the minimum distance between BSs is $500$ m.
Each BS has a maximum transmit power budget of $33$ dBm.
The path loss is $30.5 + 36.7 \log_{10}d$ in dB, where $d$ is the distance in meters.
The small scale channels follow Rayleigh fading and the noise power is $-99$ dBm.

For the proposed Edge-GNN,
all the aggregation functions are implemented by the max aggregator, which returns the element-wise maximum value of the inputs.
All the MLPs are implemented by $3$ linear layers, each followed by a ReLU activation function.
An Edge-GNN with $2$ updating layers is adopted, and the dimension of
the representations on TX-nodes, RX-nodes, and edges is set to $64$.
In the training procedure, the number of epochs is set to $500$, where
each epoch consists of $100$ mini-batches of training samples with a batch size of $256$.
For each training sample, the BSs' and UEs' locations, and the small scale channels are randomly generated.
A learning rate $\gamma=10^{-4}$ is adopted to update the trainable parameters of Edge-GNN 
by maximizing \eqref{equation-sumrate phi loss} using RMSProp \cite{tieleman2012lecture} in an unsupervised way.
During the training procedure, we set the wireless network with $5$ BSs and $2$ UEs,
while after training,
the numbers of BSs and UEs in the test procedure are set to be
larger than those in the training samples.

After training, we test the average performance of $100$ samples.
All the experiments are implemented using Pytorch on one NVIDIA V100 GPU ($32$ GB, SMX$2$).
For performance comparison, we include GP~\cite{bertsekas1997nonlinear}, which is a computationally efficient first-order algorithm for solving simply constrained optimization problems,
and WMMSE \cite{shi2011iteratively}, which in general provides high-quality solutions
to the beamforming design problem.

\begin{figure}[t]
\centering
\subfloat[Sum rate comparison]{\includegraphics[width=2.5in]{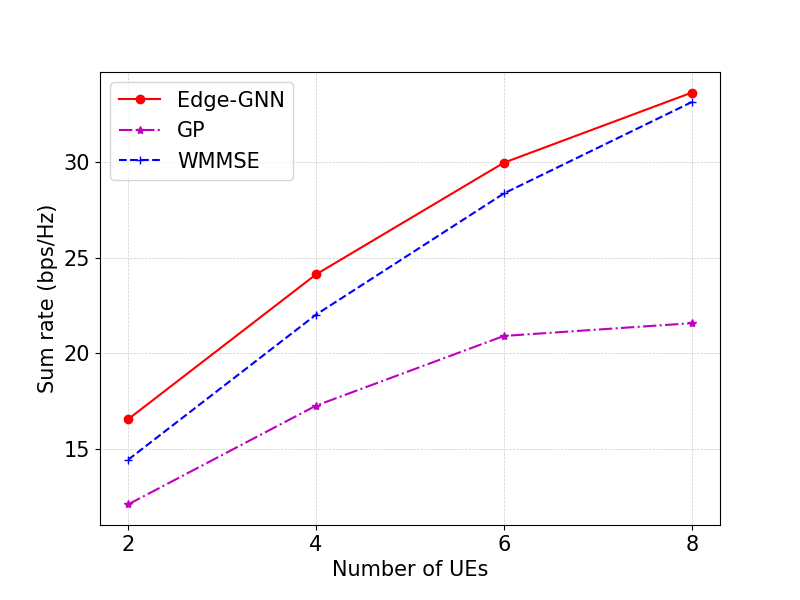}%
\label{fig-exp3 RX sr}}
\hfil
\subfloat[Computation time comparison]{\includegraphics[width=2.5in]{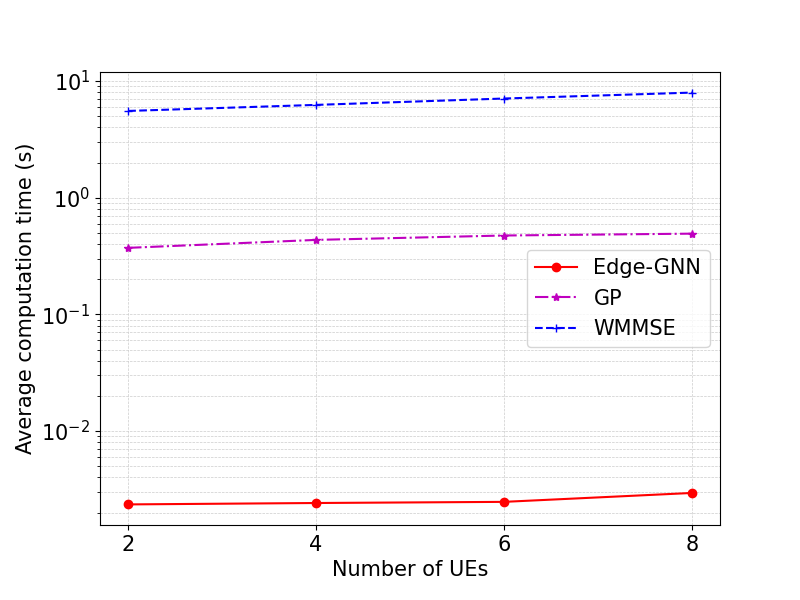}%
\label{fig-exp3 RX time}}
\caption{Generalization on number of UEs for cooperative beamforming.}
\label{fig-exp3 RX}
\end{figure}

\subsection{Generalization on Number of UEs}
To demonstrate the generalization ability of Edge-GNN with respect to different numbers of UEs,
during the training procedure, the number of UEs is fixed as $2$,
while we test the performance of the trained Edge-GNN by varying the number of UEs from $2$ to $8$.
The performance comparison in terms of sum rate and computation time is shown in Fig.~\ref{fig-exp3 RX}.
We observe from Fig.~\ref{fig-exp3 RX}\subref{fig-exp3 RX sr}
that as the number of UEs increases, Edge-GNN always
outperforms GP and WMMSE in terms of sum rate, which
demonstrates its generalization ability with respect to different numbers of UEs.
On the other hand, Fig.~\ref{fig-exp3 RX}\subref{fig-exp3 RX time} shows that Edge-GNN achieves a remarkable running speed, with over $100$ times faster than that of GP and over $1000$ times faster than that of WMMSE due to the computationally efficient feed forward computations.

\begin{figure}[t]
\centering
\subfloat[Sum rate comparison]{\includegraphics[width=1.65in]{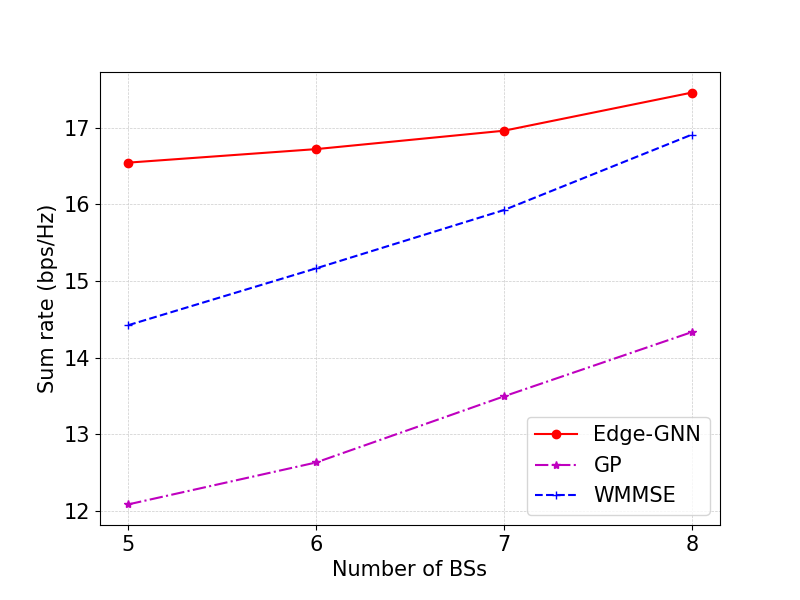}
\label{fig-TX sr}}
\hfil
\subfloat[Computation time comparison]{\includegraphics[width=1.65in]{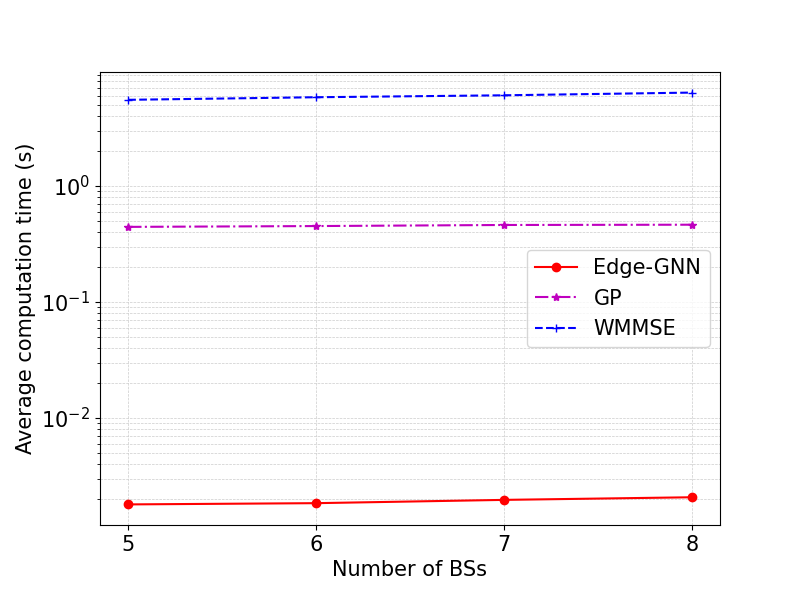}
\label{fig-TX time}}
\caption{Generalization on number of BSs for cooperative beamforming.}
\label{fig-exp1 TX}
\end{figure}

\subsection{Generalization on Number of BSs}
We further demonstrate the generalization ability of Edge-GNN with respect to different numbers of BSs.
Specifically, the number of BSs is fixed as $5$ during the training procedure,
while we test the performance of the trained Edge-GNN by varying the number of BSs from $5$ to $8$.
The performance comparison is shown in Fig.~\ref{fig-exp1 TX}.
We observe from Fig.~\ref{fig-exp1 TX}\subref{fig-TX sr} that Edge-GNN achieves higher sum rate than those of GP and WMMSE under different numbers of BSs.
Moreover, Fig.~\ref{fig-exp1 TX}\subref{fig-TX time} shows that
Edge-GNN achieves a much faster running speed
than that of GP and
WMMSE under different numbers of BSs.

\section{Conclusions}
In this paper, we investigated the edge-update empowered GNNs
to learn the cooperative beamforming in wireless networks.
The proposed edge-update mechanism
amplifies the modeling capability of GNNs, which enables
the update of variables that are defined on the edges.
Thanks to the edge-updated mechanism, the proposed Edge-GNN enjoys the PE property with respect
to both BSs and UEs, and hence achieves superior generalization ability for real-time deployment.
Simulation results demonstrated that the
proposed Edge-GNN generalizes well on different numbers of BSs and UEs, and
achieves higher sum rate with much shorter computation time than state-of-the-art
methods.

\appendices

\section{Proof of Property 2}
\label{appendix-property 2}
Let $m'=\pi_{1}(m)$ and $k'=\pi_{2}(k)$. Substituting
these two equations into \eqref{equation-EGNN encoding edge},
we have
\begin{equation}  \label{equation-PE-edge2}
\begin{aligned}
    \mathbf{e}_{m',k'}^{(l)} =~&\text{MLP}_7^{(l)}\left( \mathbf{e}_{m',k'}^{(l-1)} , \text{AGG}_{\text{E}}^{(l)}  \Big\{ \right. \\
     &\text{MLP}_5^{(l)}\left(\mathbf{e}_{m',k_1}^{(l-1)},  \mathbf{f}_{\text{BS},m'}^{(l-1)}\right), \\ &\text{MLP}_6^{(l)}\left(\mathbf{e}_{m_1,k'}^{(l-1)}, \mathbf{f}_{\text{UE},k'}^{(l-1)}   \right)
     \\ &\left. \Big\}_{k_1 \in \mathcal{N}^{\text{BS}}_{m'} \setminus \left\{k' \right\}, m_1 \in \mathcal{N}^{\text{UE}}_{k'} \setminus \left\{m'\right\}}\right),
\end{aligned}
\end{equation}
which implies that for any $\pi_1(\cdot)$ and $\pi_2(\cdot)$,
we always have \eqref{equation-PE-edge}.

\section{Proof of Proposition 1}
\label{appendix-theorem}
Since the preprocessing layer transforms the inputs $\left(\mathbf{f}_{\text{BS}}, \mathbf{f}_{\text{UE}}, \mathbf{E}\right)$
and $\left(\mathbf{\dot{f}}_{\text{BS}}, \mathbf{\dot{f}}_{\text{UE}}, \mathbf{\dot{E}}\right)$
using node/edge-wise MLPs, respectively,
the corresponding outputs satisfy
\begin{eqnarray}
\left(\mathbf{\dot{f}}_{\text{BS},\pi_1(m)}^{(0)}, \mathbf{\dot{f}}_{\text{UE},\pi_2(k)}^{(0)},  \mathbf{\dot{e}}^{(0)}_{(\pi_1(m),\pi_2(k))}\right)=
\nonumber\\
\left(\mathbf{f}_{\text{BS},m}^{(0)}, \mathbf{f}_{\text{UE},k}^{(0)},  \mathbf{e}^{(0)}_{(m,k)}\right),~~\forall m\in\mathcal{M},~\forall k\in\mathcal{K}.
\end{eqnarray}
Then, based on the PE property of the node- and edge-update mechanisms
in \eqref{PE-BS}, \eqref{PE-UE}, and \eqref{equation-PE-edge},
we further have
\begin{eqnarray}
\left(\mathbf{\dot{f}}_{\text{BS},\pi_1(m)}^{(L)}, \mathbf{\dot{f}}_{\text{UE},\pi_2(k)}^{(L)},  \mathbf{\dot{e}}^{(L)}_{(\pi_1(m),\pi_2(k))}\right)=
\nonumber\\
\left(\mathbf{f}_{\text{BS},m}^{(L)}, \mathbf{f}_{\text{UE},k}^{(L)},  \mathbf{e}^{(L)}_{(m,k)}\right),~~\forall m\in\mathcal{M},~\forall k\in\mathcal{K}.
\end{eqnarray}
Finally, since the postprocessing layer transforms $\mathbf{E}^{(L)}$
and $\mathbf{\dot{E}}^{(L)}$ using the edge-wise MLPs, respectively,
the final outputs also satisfy
$\mathbf{\dot{V}}_{(\pi_1(m),\pi_2(k),:)}=
\mathbf{V}_{(m,k,:)}$, $\forall (m,k)\in\mathcal{E}$.

\bibliographystyle{IEEEtran}
\bibliography{egbib}

\begin{thebibliography}{10}
\providecommand{\url}[1]{#1}
\csname url@samestyle\endcsname
\providecommand{\newblock}{\relax}
\providecommand{\bibinfo}[2]{#2}
\providecommand{\BIBentrySTDinterwordspacing}{\spaceskip=0pt\relax}
\providecommand{\BIBentryALTinterwordstretchfactor}{4}
\providecommand{\BIBentryALTinterwordspacing}{\spaceskip=\fontdimen2\font plus
\BIBentryALTinterwordstretchfactor\fontdimen3\font minus
  \fontdimen4\font\relax}
\providecommand{\BIBforeignlanguage}[2]{{%
\expandafter\ifx\csname l@#1\endcsname\relax
\typeout{** WARNING: IEEEtran.bst: No hyphenation pattern has been}%
\typeout{** loaded for the language `#1'. Using the pattern for}%
\typeout{** the default language instead.}%
\else
\language=\csname l@#1\endcsname
\fi
#2}}
\providecommand{\BIBdecl}{\relax}
\BIBdecl

\bibitem{zhang2004cochannel}
H.~Zhang and H.~Dai, ``Cochannel interference mitigation and cooperative
  processing in downlink multicell multiuser {MIMO} networks,'' \emph{EURASIP
  Journal on Wireless Communications and Networking}, vol. 2004, no.~2, pp.
  1--14, 2004.

\bibitem{shi2015large}
Y.~Shi, J.~Zhang, K.~B. Letaief, B.~Bai, and W.~Chen, ``Large-scale convex
  optimization for ultra-dense cloud-{RAN},'' \emph{IEEE Wireless
  Communications}, vol.~22, no.~3, pp. 84--91, 2015.

\bibitem{he2021cell}
H.~He, X.~Yu, J.~Zhang, S.~Song, and K.~B. Letaief, ``Cell-free massive {MIMO}
  for {6G} wireless communication networks,'' \emph{Journal of Communications
  and Information Networks}, vol.~6, no.~4, pp. 321--335, 2021.

\bibitem{bertsekas1997nonlinear}
D.~P. Bertsekas, ``Nonlinear programming,'' \emph{Journal of the Operational
  Research Society}, vol.~48, no.~3, pp. 334--334, 1997.

\bibitem{shi2011iteratively}
Q.~Shi, M.~Razaviyayn, Z.-Q. Luo, and C.~He, ``An iteratively weighted {MMSE}
  approach to distributed sum-utility maximization for a {MIMO} interfering
  broadcast channel,'' \emph{IEEE Transactions on Signal Processing}, vol.~59,
  no.~9, pp. 4331--4340, 2011.

\bibitem{sun2018learning}
H.~Sun, X.~Chen, Q.~Shi, M.~Hong, X.~Fu, and N.~D. Sidiropoulos, ``Learning to
  optimize: Training deep neural networks for interference management,''
  \emph{IEEE Transactions on Signal Processing}, vol.~66, no.~20, pp.
  5438--5453, 2018.

\bibitem{zhu2020learning}
M.~Zhu, T.-H. Chang, and M.~Hong, ``Learning to beamform in heterogeneous
  massive {MIMO} networks,'' \emph{arXiv preprint arXiv:2011.03971}, 2020.

\bibitem{shen2020graph}
Y.~Shen, Y.~Shi, J.~Zhang, and K.~B. Letaief, ``Graph neural networks for
  scalable radio resource management: {Architecture} design and theoretical
  analysis,'' \emph{IEEE Journal on Selected Areas in Communications}, vol.~39,
  no.~1, pp. 101--115, 2020.

\bibitem{guo2021learning}
J.~Guo and C.~Yang, ``Learning power allocation for multi-cell-multi-user
  systems with heterogeneous graph neural network,'' \emph{IEEE Transactions on
  Wireless Communications}, vol.~21, no.~2, pp. 884--897, 2021.

\bibitem{shen2022graph}
Y.~Shen, J.~Zhang, S.~Song, and K.~B. Letaief, ``Graph neural networks for
  wireless communications: {From} theory to practice,'' \emph{arXiv preprint
  arXiv:2203.10800}, 2022.

\bibitem{shen2021neural}
Y.~Shen, J.~Zhang, and K.~B. Letaief, ``How neural architectures affect deep
  learning for communication networks?'' in \emph{IEEE ICC}, 2022.

\bibitem{shen2021ai}
Y.~Shen, J.~Zhang, S.~Song, and K.~B. Letaief, ``{AI} empowered resource
  management for future wireless networks,'' in \emph{IEEE MeditCom}, 2021.

\bibitem{eisen2020optimal}
M.~Eisen and A.~Ribeiro, ``Optimal wireless resource allocation with random
  edge graph neural networks,'' \emph{IEEE Transactions on Signal Processing},
  vol.~68, pp. 2977--2991, 2020.

\bibitem{li2021heterogeneous}
Y.~Li, Z.~Chen, Y.~Wang, C.~Yang, B.~Ai, and Y.-C. Wu, ``Heterogeneous
  transformer: {A} scale adaptable neural network architecture for device
  activity detection,'' \emph{IEEE Transactions on Wireless Communications
  (accepted)}, 2022.

\bibitem{tieleman2012lecture}
T.~Tieleman, G.~Hinton \emph{et~al.}, ``Lecture 6.5-{RMSP}rop: divide the
  gradient by a running average of its recent magnitude,'' \emph{COURSERA:
  Neural networks for machine learning}, vol.~4, no.~2, pp. 26--31, 2012.

\end{thebibliography}

\end{document}